\documentclass[11pt]{article}
\usepackage{moriond,epsfig}

\def\Journal#1#2#3#4{{#1} {\bf #2}, #3 (#4)}

\def\PLB{{\em Phys. Lett.}  B}

\def\ZPC{{\em Z. Phys.} C}
\def\CPC{{\em Comp. Phys. Communications}}

\def\be{\begin{equation}}
\def\ee{\end{equation}}
\def\bea{\begin{eqnarray}}
\def\eea{\end{eqnarray}}

\newcommand{\mm}[1]{{\mbox{\hspace{#1mm}}}}
\newcommand{\beq}[1]{\begin{equation}{\label{#1}}}
\newcommand{\eeq}[0]{\end{equation}}

\newcommand{\Fig}[1]{Figure~(\ref{#1})}
\newcommand{\fig}[1]{Fig.(\ref{#1})}
\newcommand{\vev}[1]{\langle #1 \rangle}

\begin{document}
\vspace*{4cm}
\title{MINT - A SIMPLE MODEL FOR LOW ENERGY HADRONIC INTERACTIONS}

\author{M. SCHMELLING}

\address{MPI for Nuclear Physics, Saupfercheckweg 1\\
D-69117 Heidelberg, Germany}

\maketitle\abstracts{
The bulk of inelastic hadronic interactions is
characterized by longitudinal phase space and exponentially
damped transverse momentum spectra. A simple model with only 
a single adjustable parameter is presented, making it a very 
convenient tool for systematic studies, which gives a surprisingly 
good description of $pA$-collisions at 920 GeV beam energy.}

\section{Introduction}
\begin{figure}[ht]
\centerline{%
\epsfig{figure=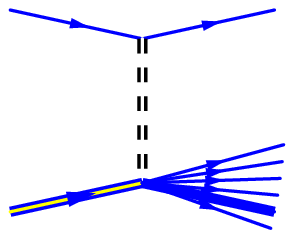,width=0.3\textwidth,height=32.0mm}\mm{16}%
\epsfig{figure=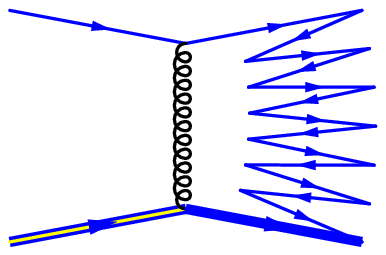,width=0.3\textwidth,height=32.8mm}}
\caption{Examples for inelastic $pA$\/ collisions showing diffractive 
         scattering (left hand side), mediated without colour exchange 
         by a pomeron, in comparison to a normal inelastic interaction 
         (right) involving gluon exchange.}
\label{fig:minbias}
\end{figure}

As illustrated in \fig{fig:minbias}, proton-nucleus ($pA$) collisions
can proceed with or without colour exchange. The latter case 
applies for example in diffractive scattering, where beam or target 
can emerge in a high mass excited state. Reactions with colour 
exchange, referred to as ``normal inelastic interactions'' in the following, 
constitute the bulk of the total cross section. Here a single high 
mass system is created which decays into the observable final state particles. 

Interactions without colour exchange are characterized by an 
exponential spectrum $dn/dt \sim e^{bt}$\/ for the momentum transfer 
$t$\/ and mass spectra $dn/dM\sim 1/M$\/ for diffractively excited 
states, essentially independent of the target nucleus $A$. This is 
consistent with the picture that reactions without colour exchange are 
peripheral, involving momentum transfer only between the beam proton 
and a single nucleon of the target nucleus. In contrast, normal inelastic
interactions show a significant $A$-dependence, indicating a more central 
character of such collisions.     

A Monte Carlo model for the generation of exclusive final 
states in $pA$\/ collisions can be constructed in a straightforward way, 
using as external input the measured cross sections for the different 
sub-processes and a prescription for the transition of an initial high 
mass system into final particles. The latter is provided by MINT, based
on the assumption that this transition is universal, i.e. it is the 
same for normal inelastic interactions or diffractive masses.

\section{The MINT Model}
The phenomenological basis of the MINT model is the observation that in 
hadronic interactions with center-of-mass energy $E_{cm}$\/ large against 
the nucleon mass, the typical transverse momenta of final state particles 
are negligible compared to their longitudinal momenta. Assuming further 
that dynamical effects such as hard parton-parton scattering can be ignored, 
which is a reasonable approximation at low energies, the final state is 
governed by longitudinal phase space. Then, for zero transverse momentum, 
$p_T=0$, the Lorentz invariant phase space element of a free particle, 
$d^3p/2E =dp_T^2\;d\phi\;dy/4$, implies a uniform particle density 
in rapidity $y$. In MINT, for normal inelastic interactions the longitudinal 
direction in the center-of-mass system is along the axis defined by the 
momenta of the colliding particles. For diffractive scattering it is 
chosen along the direction of the outgoing systems.

To reconcile longitudinal phase space and finite transverse momenta
of the final state particles, MINT employs a two step procedure. 
The basic idea is to start by generating primary clusters with zero 
transverse momentum, which then perform two-body decays into the actual
final state particles. The mass spectrum of the primary clusters 
\beq{dndm}
   \frac{dn}{dm} = a^2 m e^{-am} 
   \mm{10}\mbox{easily generated by}\mm{10}
    m = -\frac{1}{a}\ln(r_1 r_2)
\eeq 
from two uniform random deviates $r_1,r_2 \in [0,1]$\/ and the still to be 
defined parameter $a$, thus determines the $p_T$-spectrum of the final 
state particles. The case of massless secondaries can be solved analytically 
to yield $dn/dp_T = 4 a^2 p_T K_0(2ap_T)$\/ where $K_0$\/ is the modified Bessel 
function of order zero. For large $p_T$\/ it shows an approximately
exponential behaviour, in agreement with observations. The mean transverse 
momentum is given by $\vev{p_T}=\pi/4a$\/ which motivates to use the rescaled 
quantity $\alpha=4a/\pi$\/ as the free parameter of the model. For decays 
into massive secondaries the spectrum will be slightly different, but will 
not change qualitatively.

Given the mass $m$\/ of a cluster, its rapidity is generated uniformly 
over the kinematically allowed range $\pm\ln(E_{cm}/m)$. The generation 
of mass and rapidity is iterated until the total invariant mass $M$\/ 
of the system exceeds $E_{cm}$. Then the generation stops and
the primary clusters are shifted in rapidity such that the longitudinal 
momentum balances, using $p_L=m_T \sinh y$\/ and $m_T=m$\/ for $p_T=0$. 
Finally all masses are scaled such that $M=E_{cm}$, i.e. MINT satisfies exact
energy-momentum conservation. Note also that in this scheme the particle
multiplicity distribution is an absolute prediction by the model.

Since flavour physics is not the objective of MINT, only decays into 
pions and photons are considered. Clusters with a mass $m<2m_{\pi}$, where 
$m_{\pi}$\/ is the charged-pion mass, are neutral and decay into two photons. 
Heavier clusters are assumed to behave like $\rho$-mesons. They are randomly
assigned charges $Q=\{-1,0,+1\}$\/ with equal probability but subject to the 
constraint of global charge conservation. Charged primaries decay according to 
$X^{\pm}\rightarrow \pi^\pm\pi^0 \rightarrow\pi^\pm \gamma\gamma$\/ and 
neutral ones via the mode $X^0\rightarrow\pi^+\pi^-$. 

The above discussion defines the generation of final state particles 
from a single high mass system. It remains to specify how normal inelastic 
$pA$\/ interactions shall be modeled. In MINT this is done as the 
incoherent sum of $n$\/ subsystems. Here $1\leq n \leq A$\/ is the number of 
participating nucleons from the target nucleus, drawn from a Poisson distribution 
with mean value given by the number of nucleons intercepted by the beam proton.
A uniform distribution of the impact point of the beam proton on the target
nucleus is assumed. Every subsystem has the same invariant mass, which is a 
natural assumption for the case where the beam proton hits the nucleus at
rest, its charge is randomly chosen from $Q=\{0,+1\}$, with a probability 
$Z/A$ for $Q=+1$. To account for the charge of the beam proton, the charge 
of the first subsystem is increased by one unit. A more detailed discussion 
of the implementation of the model can be found in a HERA-B note describing 
the MINT model \cite{ref:hbnote}.

\section{Model Tuning and Comparison with Real Data}
\begin{figure}[ht]
\begin{center}
\begin{minipage}[h]{0.55\textwidth}
\epsfig{figure=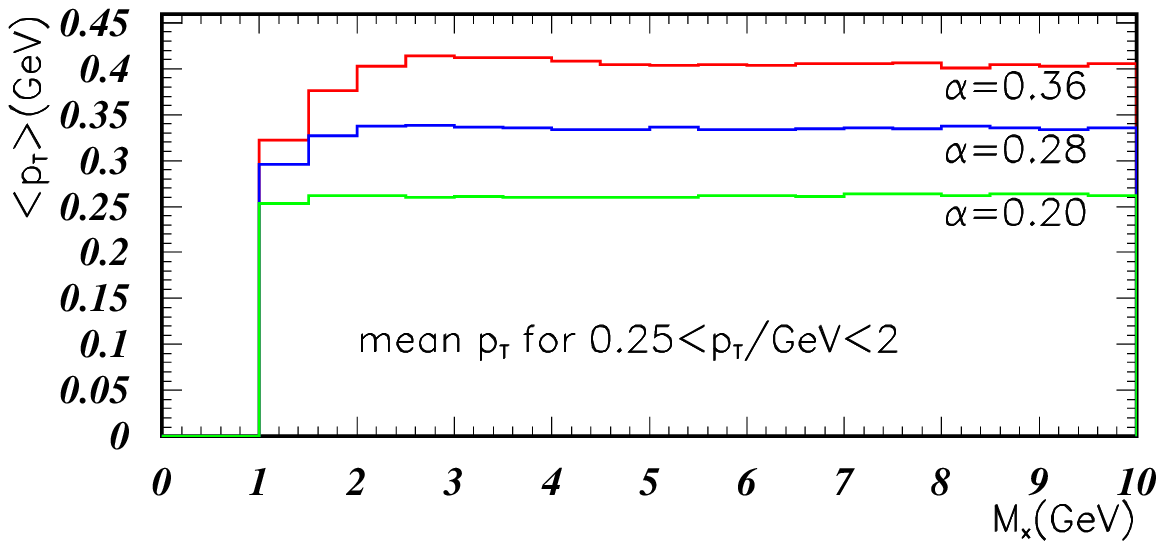,width=\textwidth}
\end{minipage}
\begin{minipage}[h]{0.37\textwidth}
\vspace*{-4mm}
\epsfig{figure=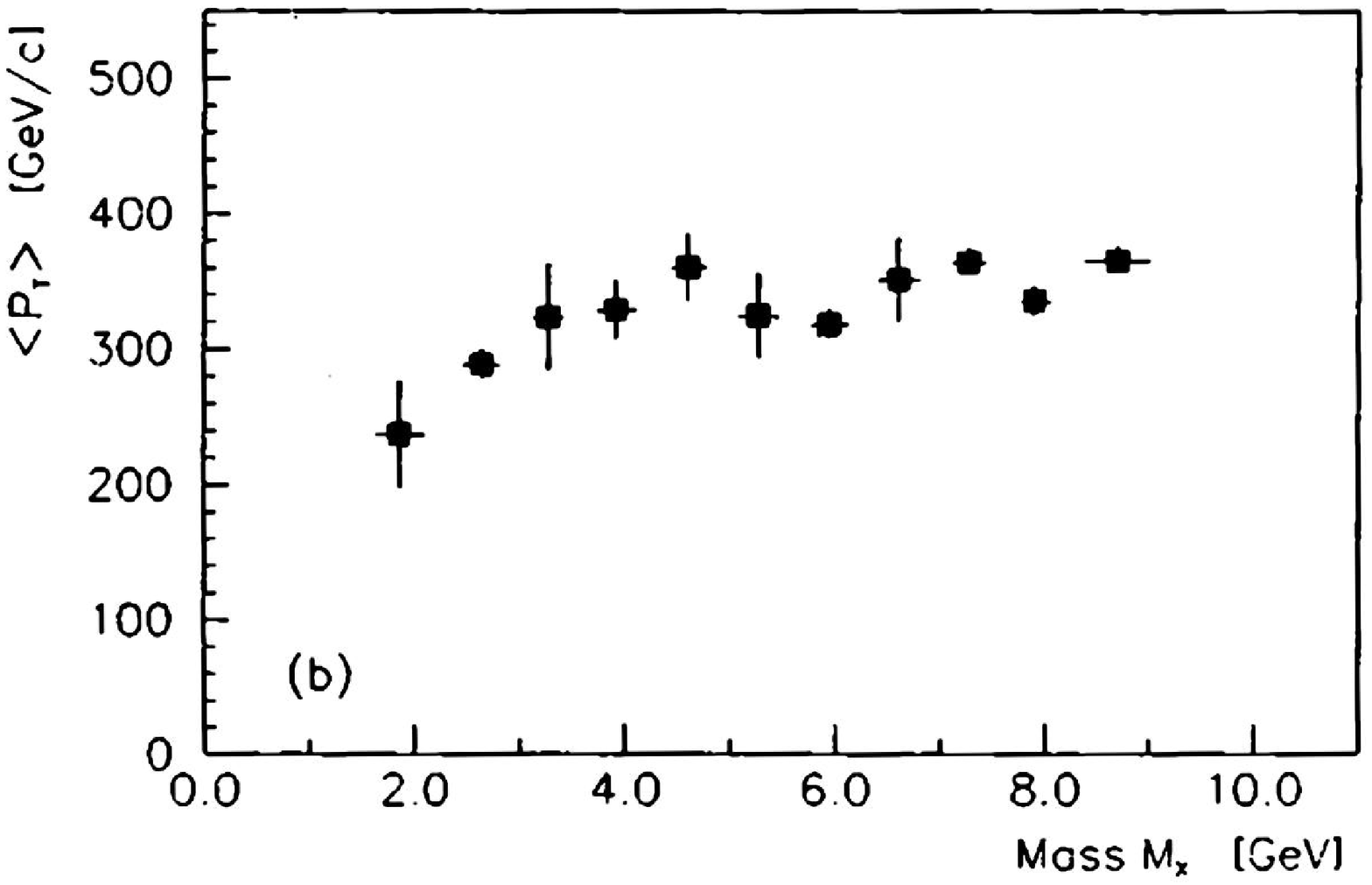,width=\textwidth}
\end{minipage}
\end{center}
\caption[]{Average transverse momentum of secondaries from target single 
           diffractive scattering
           as a function of the diffractive mass for MINT (left)
           and real data \cite{ref:Ake91} (right -- here the units of 
           the ordinate should read MeV/$c$). }
\label{fig:tuning}
\end{figure}

Since MINT is based on the assumption that the transition of a primary
high mass system into final state particles is universal, the adjustment
of the only free parameter of the model is done to the transverse momentum 
spectrum observed in target single diffraction processes~\cite{ref:Ake91}.
\Fig{fig:tuning} shows for three values $\alpha$\/ how, as a function of the 
mass $M_x$\/ of the diffractive system, MINT compares with data for the
average $p_T$\/ in the range $0.25\;\mbox{GeV}/c < p_T <2\;\mbox{GeV}/c$.
The value $\alpha=0.28$\/ roughly matches the measurements and was used 
throughout later on. Given the relative simplicity of the model, no attempt 
was made to perform an actual fit to the data. 

\begin{figure}[ht]
\begin{center}
\epsfig{figure=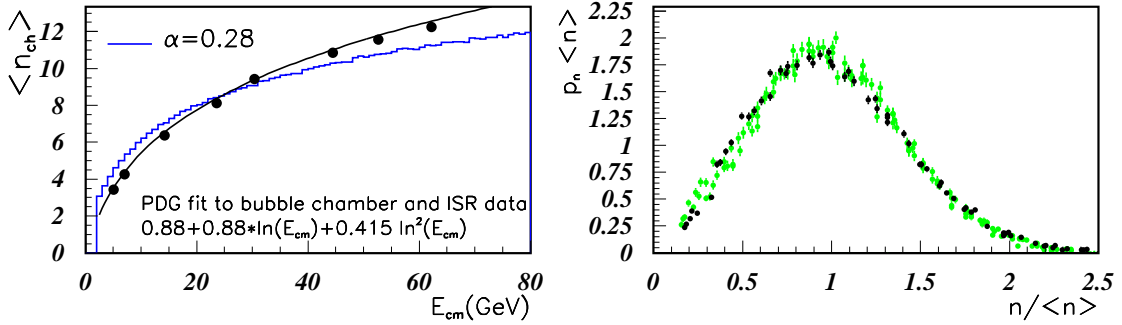,width=0.92\textwidth}
\caption[]{Multiplicities in inelastic $pp$-collisions. On the left the 
           average charged multiplicities predicted by MINT (histogram) 
           compared to a compilation of measurements (points) and a global 
           parameterization \cite{ref:pdg88} (line) by the PDG as a function
           of $E_{cm}$. The right hand side displays for $E_{cm}$ between 10
           and 60~GeV the scaled multiplicities for charged (black dots) 
           and all particles (grey dots), showing that approximate KNO 
           scaling holds in MINT.} 
\label{fig:nchpp}
\end{center}
\end{figure}

As a first test, in \fig{fig:nchpp} the mean charged particle multiplicities
for inelastic proton-proton collisions predicted by MINT as a 
function of the center-of-mass energy is compared to a real data. Up to the
energies reached at the CERN Intersecting Storage Rings (ISR) the agreement 
is surprisingly good, with a discrepancy below one track per event. 
Approximate KNO scaling is found for charged and total multiplicities 
in the energy range from $E_{cm}=10$ to $60$ GeV.

In addition, a comparison was done with data from $pA$\/ collisions 
recorded by the HERA-B~\cite{ref:herab} fixed-target experiment at the 
HERA storage ring of DESY/Hamburg. The nucleon-nucleon center-of-mass energy 
was $\sqrt{s_{NN}}=41.5$~GeV. The detector is a forward magnetic spectrometer
with an angular acceptance of $15-220$~mrad in the bending plane. The
tracking systems consists of a vertex detector (VDS) before and Outer Tracker
chambers behind the magnet; particle identification is performed by a 
RICH detector, an electromagnetic calorimeter (ECAL) and a muon system. 
Target materials used in 2002/03 were Carbon, Titanium and Tungsten.

\begin{figure}[ht]
\centerline{%
\epsfig{figure=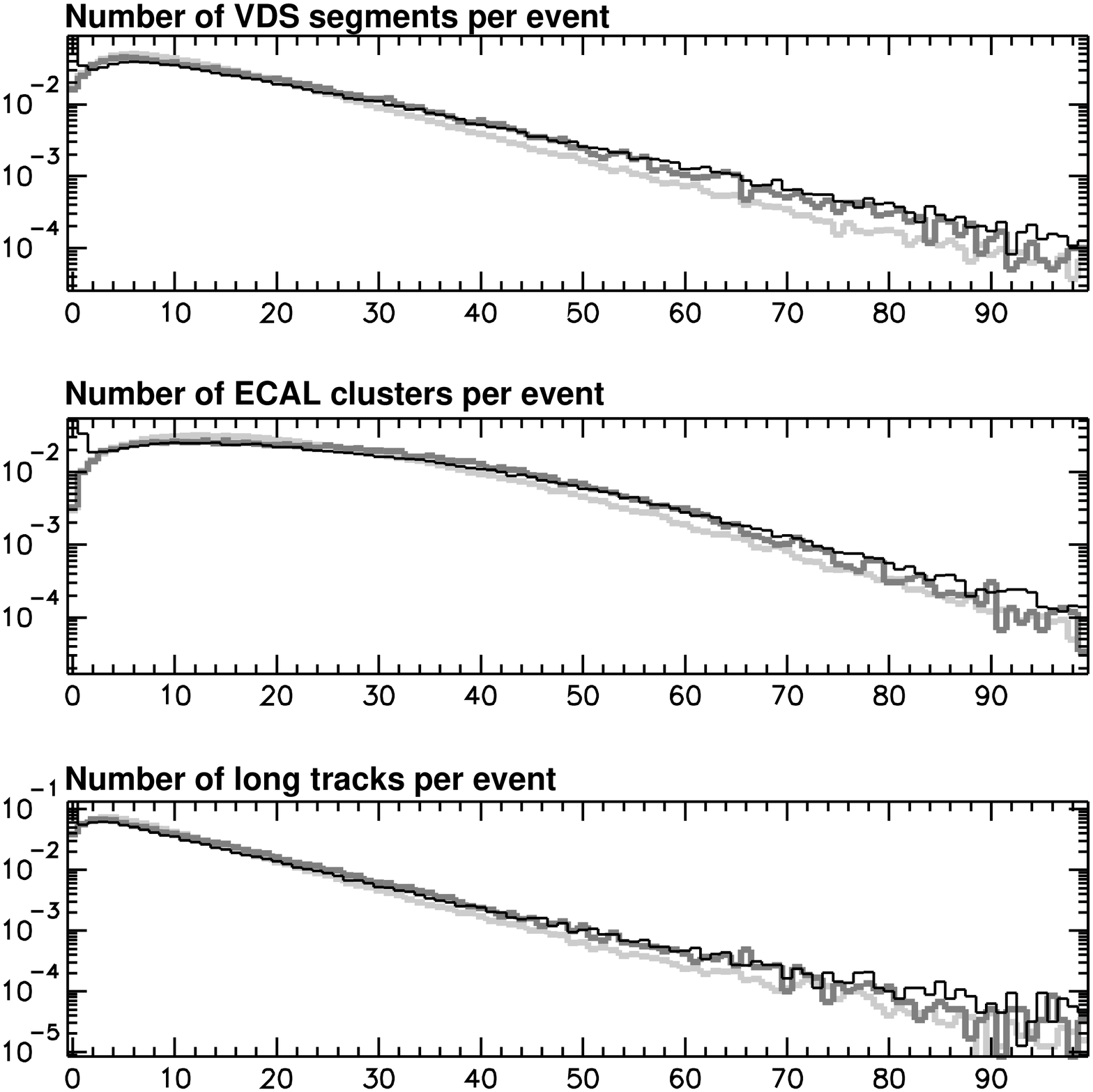,width=0.46\textwidth}\mm{4}%
\epsfig{figure=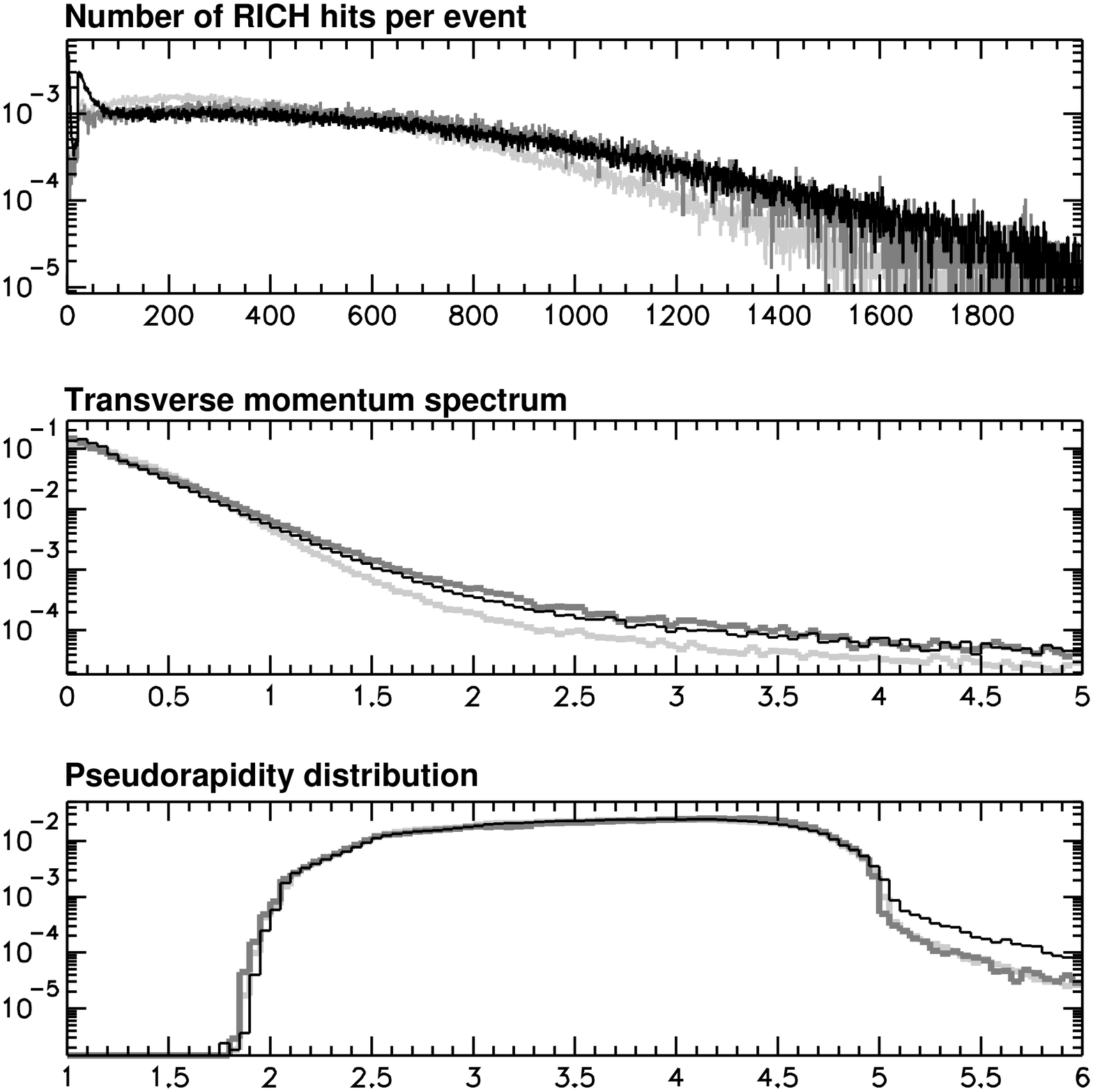,width=0.46\textwidth}}
\caption{Comparison between HERA-B raw data and Monte-Carlo models 
         after full detector simulation. Thin black lines are
         real data, the prediction by FRITIOF and MINT 
         are shown in light and dark grey, respectively.} 
\label{fig:herab}
\end{figure}

\Fig{fig:herab} shows how MINT and the FRITIOF~\cite{ref:FRITIOF} 
model after the full detector simulation compare to real 
data. The comparison covers sub-detector specific quantities, such as track 
segments reconstructed in the VDS, the number of hits seen in the RICH and 
the number of clusters per event from the ECAL, as well as physics quantities 
like the number of charged tracks per event passing through the entire 
tracking system, the transverse momentum distribution and the 
pseudo-rapidity distribution of those tracks. In general, MINT describes 
the data as well as FRITIOF. Given the simplicity and minimal amount 
of tuning that went into the model, it is surprisingly accurate. Interestingly, 
the number of hits in the RICH and the transverse momentum spectrum with its 
high-$p_T$\/ tail is better reproduced by MINT than by FRITIOF. The same qualitative 
findings apply for the lighter Carbon and the heavier Tungsten target 
without retuning the model.

\vspace*{-2mm}
\section{Summary}
A simple and surprisingly accurate model has been presented for the
description of $pA$-collisions with nucleon-nucleon center-of-mass energies 
up to $E_{cm}\sim 60$~GeV. The model has only a single adjustable parameter,
which makes it very convenient for systematic studies exploring the 
sensitivity of a physics analysis to details of a Monte Carlo model. 
MINT, which incorporates also elastic and diffractive scattering in its 
implementation satisfies exact energy-momentum and charge conservation, 
but features only charged pions and photons in the final state.

\section*{Acknowledgments}
Sincere thanks go to Mikhail Zavertyaev and Marco Bruschi for the detailed
comparison between MINT and HERA-B data.

\section*{References}

\end{document}